\documentclass{AIMS}
\usepackage{amsmath}
\usepackage{epsfig} 
\usepackage{graphics} 

\def\currentvolume{2}
 \def\currentissue{2}
  \def\currentyear{2007}
   \def\currentmonth{June}
    \def\ppages{193--210}

\theoremstyle{definition}

\title[Self-Organized Network Flows]
      {Self-Organized Network Flows}

\author[Dirk Helbing, Jan Siegmeier, and Stefan L\"ammer]{}


\email{helbing@trafficforum.org}

\begin{document}
\maketitle

\centerline{\scshape Dirk Helbing,$^{1,2}$ Jan Siegmeier,$^1$ and Stefan L\"ammer$^1$}
\medskip
{\footnotesize
 \centerline{$^1$ Institute for Transport \& Economics, Dresden University of Technology}
\centerline{Andreas-Schubert-Str. 23, 01062 Dresden, Germany} \centerline{$^2$ Collegium Budapest~-- Institute for Advance Study}
   \centerline{Szenth\'{a}roms\'{a}g utca 2, 1014 Budapest, Hungary}
}

\medskip

 \centerline{(Communicated by Aim Sciences)}
 \medskip

\begin{abstract}
A model for traffic flow in street networks or material flows in supply networks is presented, that takes into account the conservation of
cars or materials and other significant features of traffic flows such as jam formation, spillovers, and load-dependent transportation times.
Furthermore, conflicts or coordination problems of intersecting or merging flows are considered as well. Making assumptions regarding
the permeability of the intersection as a function of the conflicting flows and the queue lengths, we find self-organized oscillations in the
flows similar to the operation of traffic lights.
\end{abstract}

\section{Introduction}

Material flows are found in many places of the world. This concerns, for example, traffic flows in urban areas or flows of commodities in
logistic systems. There is also some similarity with material flows in production or biological systems, from cells over bodies upto
ecological food chains. Many of these material flows are not of diffusive nature or going on in continuous space. They are often directed and
organized in networks. In comparison with data flows in information networks, however, there are conservation laws, which can be used to set
up equations for material flows in networks. It turns out, however, that this is not a trivial task. While there is already a controversial
discussion about the correct equations representing traffic flows along road sections \cite{Daganzo,RMP,mitSchoenhof,Kernerbuch}, their
combination in often complex and irregular networks poses further challenges. In particular, there have been several publications on the
treatment of the boundary conditions at nodes (connections) of several network links (i.e. road sections)
\cite{Hilliges,CellTransmission,Lebacque,control,Klar,Piccoli1,Piccoli2,Herty3,Herty2,Herty1}.
In particular, the modelling of merging and intersecting flows
is not unique, as there are many possible forms of organization, including the use of traffic lights. Then, however, the question comes up
how these traffic lights should be operated, coordinated, and optimized. In order to address these questions, in Sec.~\ref{Sec2} we formulate
a simple model for network flows, which contains the main ingredients of material or traffic flows. Section \ref{Sec3} will then discuss the
treatment of diverges, merges, and intersections. Equations for the interaction-dependent permeability at merging zones and intersections
will be formulated in Sec.~\ref{Sec4}. We will see that, under certain conditions, they lead to spontaneous oscillations, which have features
similar to the operation of traffic lights. Finally, Sec.~\ref{Sec5} summarizes and concludes this paper.

\section{Flows in Networks}\label{Sec2}

The following section will start with a summary of the equations derived for traffic flows in networks in a previous paper. These equations
are based on the following assumptions:
\begin{itemize}
\item The road network can be decomposed into road sections of homogeneous capacity
(links) and nodes describing their connections.
\item The traffic dynamics along the links is sufficiently well described by the
Lighthill-Whitham model, i.e. the continuity equation for vehicle conservation and a flow-density relationship (``fundamental diagram'').
This assumes adiabatic speed adjustments, i.e. that acceleration and deceleration times can be neglected.
\item The parameters of vehicles such as the maximum speed $V_i^0$ and the safe time headway $T$
are assumed to be identical in the same road section, and who enters a road section first exits first (FIFO principle). That is, overtaking
is assumed to be negligible.
\item The fundamental diagram can be well approximated by a triangular shape, with an
increasing slope $V_i^0$ at low densities and a decreasing slope $c$ in the congested regime. This implies two constant characteristic
speeds: While $V_i^0$ corresponds to the free speed or speed limit on road section $i$,
\begin{equation}
 - c = -\frac{1}{\rho_{\rm max} T}
\end{equation}
the dissolution speed of the downstream front of a traffic jam and the velocity of upstream propagation of perturbations in congested traffic.
While $\rho_{\rm max}$ denotes the maximum vehicle density in vehicle queues, $T\approx 1.8$s is the safe time gap between two successive
vehicles.
\item The vehicle density in traffic jams is basically constant.
\end{itemize}
These assumptions may be compensated for by suitable corrections \cite{RMP}, but already the model below displays a rich spectrum of
spatio-temporal behaviors and contains the main elements of traffic dynamics we are interested in here.

\subsection{Flow Conservation Laws}\label{laws}

In the following, we will introduce our equations for traffic flows in networks very shortly, as a detailed justification and derivation has
been given elsewhere \cite{JPhysA,TGF03,control}. These equations are also meaningful for pipeline networks \cite{pipelines} (if
complemented by equations for momentum conservation), logistic
systems \cite{logistics}, or supply networks \cite{supply}. Our notation is illustrated in Fig.~\ref{Fig1}.
\par
Compared to Ref.~\cite{control}, we will use a simplified notation,
here.\footnote{The arrival flow $A_j(t)$ has previously
been denoted by  $Q_j^{\rm arr}(t)$, the potential arrival flow $\widehat{A}_j(t)$ by $Q_j^{\rm arr, pot}(t)$,
the departure flow $O_j(t)$ by $Q_j^{\rm dep}(t)$ and the potential departure flow $\widehat{O}_j(t)$ by
$Q_j^{\rm dep,pot}(t)$.} The {\em arrival flow} $A_j(t)$ denotes the actual inflow of vehicles into the upstream
end of road section $j$, while $O_j(t)$ is the actual {\em departure flow}, i.e. the flow of vehicles
leaving road section $j$ at its downstream end. The quantity
\begin{equation}
 \widehat{Q}_j = \left( T + \frac{1}{V_j^0\rho_{\rm max}} \right)^{-1}
 = \frac{\rho_{\rm max}}{1/c + 1/V_j^0}
\end{equation}
represents the maximum in- or outflow of road section $j$.
All the above quantities refer to flows {\it per lane}.
$I_j$ is the number of lanes and $L_j$ the length of road section $j$.
$l_j(t)\le L_j$ is the length of the congested area on link $j$ (measured from the
downstream end),
and $\Delta N_j$ is the number of stopped or delayed vehicles, see Eqs. (\ref{shock}) and (\ref{delayed}).
With these definitions, we can formulate constraints for the {\em actual} arrival and departure flows,
which are given by the {\em potential arrival flows} $\widehat{A}_j(t)$ and the {\em potential
departure flows} $\widehat{O}_i(t)$, respectively.
\par
The actual arrival flow $A_j(t)$ is limited by the maximum inflow $\widehat{Q}_j$, if
road section $j$ is not fully congested ($l_j(t) < L_j$). Otherwise (if $l_j=L_j$) it is limited
by the actual departure flow $O_j(t-L_j/c)$ a time period $L_j/c$ before,
as it requires this time period until the downstream flow value has propagated upto the upstream end
of the road section by forward movement of vehicles under congested traffic conditions.
This implies
\begin{equation}
 0 \le A_j(t) \le \widehat{A}_j(t) := \left\{
\begin{array}{ll}
\widehat{Q}_j & \mbox{if } l_j(t) < L_j \\
O_j(t-L_j/c) & \mbox{if } l_j(t) = L_j.
\end{array}
\right. \label{one}
\end{equation}
Moreover, the potential departure flow $\widehat{O}_i(t)$ of road section $i$
is given by its permeability $\gamma_i(t)$ times the
maximum outflow $\widehat{Q}_i$ from this road section,
if vehicles are queued up ($\Delta N_i >0$) and waiting to leave.
Otherwise (if $\Delta N_i =0$) the outflow is limited by the permeability times the
arrival flow $A_i$ a time period $L_i/V_i^0$ before,
as this is the time period that entering vehicles need to reach the end of road section $i$
when moving freely at the speed $V_i^0$.
This gives the additional relationship
\begin{equation}
0 \le O_i(t) \le \widehat{O}_i(t) := \gamma_i(t) \left\{
\begin{array}{ll}
A_i(t-L_i/V_i^0) & \mbox{if } \Delta N_i(t) = 0 \\
\widehat{Q}_i & \mbox{if } \Delta N_i(t) > 0 \, .
\end{array}
\right. \label{two}
\end{equation}
The permeability $\gamma_i(t)$ for traffic flows at the downstream end of section $i$
can assume values between 0 and 1. In case of a traffic light, $\gamma_i(t) = 1$ corresponds
to a green light for road section $i$, while $\gamma_i(t) =0$ corresponds to a red or amber
light.
\par
Alternatively and shorter than Eqs.~(\ref{one}) and (\ref{two}) one can write
\begin{equation}
 \widehat{A}_j(t) = \max\Big[ \widehat{Q}_j \Theta(l_j(t) < L_j),O_j(t-L_j/c)\Big]
\label{easy1}
\end{equation}
and
\begin{equation}
 \widehat{O}_i(t) = \gamma_i(t) \max \Big[\widehat{Q}_i\Theta(\Delta N_i > 0), A_i(t-L_i/V_i^0)\Big] \, ,
\label{easy2}
\end{equation}
where the Heaviside function $\Theta$ is 1, if the argument (inequality) has the logical value ``true'', otherwise it is 0. Note
that the above treatment of the traffic flow in a road section requires the specification of the boundary conditions only, as we have
integrated up Lighthill's and Whitham's partial differential equation over the length of the road section. The dynamics in the inner part of
the section can be easily reconstructed from the boundary conditions thanks to the constant characteristic speeds. However, a certain point
of the road section may be determined either from the upstream boundary (in the case of free traffic) or by the downstream boundary (if lying
in the congested area, i.e. behind the upstream congestion front). Therefore, we have a switching between the influence of the upstream and
the downstream boundary conditions, which makes the dynamics both, complicated and interesting. This switching results from the maximum
functions above and implies also that material flows in networks are described by hybrid equations. Although the dynamics is determined by
linear ordinary differential equations in all regimes, the switching between the regimes can imply a complex dynamics and even deterministic
chaos \cite{Peters}.
\par
Complementary to the above equations, we have now to specify the constraints for the nodes, i.e. the connection, merging, diverging or
intersection points of the homogeneous road sections. Let the ingoing links be denoted by the index $i$ and the outgoing ones by $j$. To
distinguish quantities more easily when we insert concrete values $1,2,\dots$ for $i$ and $j$, we mark quantities of outgoing links
additionally by a prime (${}'$).
\par
Due to the condition of flow conservation, the arrival flow into a road section $j$ with $I'_j$ lanes must agree with the sum of the
fractions $\alpha_{ij}$ of all outflows $I_iO_i(t)$ turning into road section $i$. Additionally, the arrival flows are limited, i.e. we have
\begin{equation}
 I'_jA'_j(t) = \sum_i I_iO_i(t) \alpha_{ij} \le I'_j\widehat{A}'_j(t)
\label{inequal}
\end{equation}
for all $j$. Of course, the turning fractions $\alpha_{ij}\ge 0$ are normalized due to flow conservation:
\begin{equation}
 \sum_j \alpha_{ij}(t) = 1 \, .
\end{equation}
\par
In cases of no merging flows, Eq.~(\ref{inequal}) simplifies to
\begin{equation}
 I'_jA'_j(t) = I_iO_i(t) \alpha_{ij} \le I'_j\widehat{A}'_j(t)
\end{equation}
for all $j$. At the same time, $0 \le O_i(t) \le \widehat{O}_i(t)$ must be fulfilled for all $i$. Together, this implies
\begin{equation}
O_i(t) \le \min \left[ \widehat{O}_i(t) , \min_j \left( \frac{I'_j \widehat{A}'_j}{I_i\alpha_{ij}} \right) \right] \label{nomerge}
\end{equation}
for all $i$.
\par
The advantage of the above model is that it contains the most important elements of the traffic dynamics in networks. This includes the
transition from free to congested traffic flows due to lack of capacity, the propagation speeds of vehicles and congested traffic, spillover
effects (i.e. obstructions when entering fully congested road sections) and, implicitly, load-dependent travel times as well.

\subsection{Two Views on Traffic Jams}

Let us study the traffic dynamics on the road sections in more detail.
Traffic  jams can be handled in two different ways: First by
determining the number of cars that are delayed compared to free traffic or, second, by determining fronts and ends of traffic jams. The
former method is more simple, but it cannot deal correctly with spill-over effects, when the end of a traffic jam reaches the end of a road
section. Therefore, the first method is sufficient only in situations where the spatial capacity of road sections is never exceeded.

\subsubsection{Method 1: Number of Delayed Vehicles}

The first method just determines the difference between the number $N_i^{\rm in}(t)$ of vehicles that would reach the end of road section $i$
upto time $t$ and the number $N_i^{\rm out}(t)$ of vehicles that actually leave the road section upto this time. $N_i^{\rm in}(t)$ just
corresponds to the number of vehicles which have entered the road section upto time $t-L_i/V_i^0$, as $L_i/V_i^0$ is the free travel time.
This implies
\begin{equation}
 N_i^{\rm in}(t) = \int\limits_0^t dt' \; A_i(t'-L_i/V_i^0) \, ,
\end{equation}
while the number of vehicles that have acually left the road section upto time $t$ is
\begin{equation}
 N_i^{\rm out}(t) = \int\limits_0^t dt' \; O_i(t') \, .
\end{equation}
Hence, the number $\Delta N_i(t)$ of delayed vehicles is given by
\begin{equation}
  \Delta N_i(t) = \int\limits_0^t dt' \; [A_i(t'-L_i/V_i^0) - O_i(t')] \ge 0 \, .
\end{equation}
Alternatively, one can use the following differential equation for the temporal change in the number of delayed vehicles:
\begin{equation}
  \frac{d\,\Delta N_i}{dt} = A_i(t-L_i/V_i^0) - O_i(t) \, .
\label{delayed}
\end{equation}
In contrast, the number of {\em all} vehicles on road section $i$ (independently of whether they are delayed or not) changes in time
according to
\begin{equation}
  \frac{dN_i}{dt} = A_i(t) - O_i(t) \, .
\end{equation}

\subsubsection{Method 2: Jam Formation and Resolution}

In our simple macroscopic traffic model, the formation and resolution of traffic jams is described by the shock wave equations, where we have
the two characteristic speeds $V_i^0$ (the free speed) and $c$
(the jam resolution speed). According to the theory of shock waves \cite{LW,Whitham}, the upstream end of a traffic jam, which is located at
a place $l_i(t)\ge0$ upstream of the end of road section $i$, is moving at the speed
\begin{equation}
 \frac{dl_i}{dt} = - \frac{A_i\big(t-[L_i-l_i(t)]/V_i^0\big) - O_i\big(t-l_i(t)/c\big)
}{\rho_1(t) - \rho_2(t)}  \, \label{shock}
\end{equation}
with the (free) density
\begin{equation}
 \rho_1(t) = A_i\big(t-[L_i-l_i(t)]/V_i^0\big)/V_i^0
\end{equation}
immediately before the upstream shock front and the (congested) density
\begin{equation}
 \rho_2(t) =  [1-TO_i\big(t-l_i(t)/c\big)]\rho_{\rm max}
\end{equation}
immediately downstream of it. This is, because free traffic is upstream of the shock front, and congested traffic downstream of it (for
details see Eqs. (1.6) and (1.4) in Ref.~\cite{control}). In contrast, the downstream front of a traffic jam is moving at the speed
\begin{equation}
 - \frac{0  - O_i\big(t-l_i(t)/c\big)
}{\rho_{\rm max} - O_i\big(t-l_i(t)/c\big)/V_i^0 } = \frac{O_i\big(t-l_i(t)/c\big) }{\rho_{\rm max} - O_i\big(t-l_i(t)/c\big)/V_i^0 }
\, ,
\end{equation}
since congested traffic with zero flow is upstream of the shock front and free traffic flow occurs downstream of it.

\subsubsection{Comparison of the Two Methods}\label{compa}

Let us discuss a simple example to make the differences of both descriptions clearer. For this, we assume that, at time $t=0$, traffic flow
on the overall road section $i$ is free, i.e. any traffic jam has resolved and there are no delayed vehicles. The flow shall be stopped by a
red traffic light for a time period $t_0$. At time $t=t_0$, the traffic light shall turn green, and the formed traffic jam shall resolve. For
the arrival flow, we simply assume a constant value $A_i$, and the road section shall be long enough to take up the forming traffic jam.
Moreover, the departure flow shall be $O_i$. Then, according to method 1, the number of delayed vehicles at time $t_0$ is
\begin{equation}
 \Delta N_i(t_0) = A_i t_0 \, ,
\end{equation}
and it is reduced according to
\begin{equation}
 \Delta N_i(t) = A_i t_0 - (O_i - A_i) (t-t_0) \, .
\end{equation}
Therefore, any delays are resolved after a time period
\begin{equation}
 t-t_0  = \frac{A_it_0}{O_i - A_i} = \frac{\Delta N_i(t_0)}{O_i -A_i} \, ,
\label{dis1}
\end{equation}
i.e. at time
\begin{equation}
 t_2 = t_0 \frac{O_i}{O_i - A_i} \, .
\label{acco}
\end{equation}
Afterwards, $\Delta N_i(t) = 0$.
\par
In contrast, the end of the traffic jam grows with the speed
\begin{equation}
 \frac{dl_i}{dt} = - \frac{A_i - 0} {A_i/V_i^0 - (1-0)\rho_{\rm max}}
= \frac{1}{\rho_{\rm max}/A_i - 1/V_i^0} =: C_i \, .
\end{equation}
Therefore, we have $l_i(t_0) = C_i t_0$. Surprisingly, this is greater than $\Delta N_i(t_0)/\rho_{\rm max}$, i.e. the expected length of the
traffic jam based on the number of delayed vehicles. The reason is that the delay of a vehicle joining the traffic jam at location $x_i = L_i
- l_i$ is noticed at the downstream end of the road section only after a time period $l_i/V_i^0$.
\par
The resolution of the traffic jam starts from the downstream end with the speed
\begin{equation}
  \frac{0-\widehat{Q}_i}{\rho_{\rm max} - \widehat{Q}_i/V_i^0} = \frac{-1}{\rho_{\rm max}/\widehat{Q}_i - 1/V_i^0} = -c \, ,
\end{equation}
if the outflow is free (i.e. $O_i = \widehat{Q}_i$), otherwise with the speed
\begin{equation}
 \frac{0-O_i}{\rho_{\rm max} - (\rho_{\rm max} - O_i/c)} = - c \, ,
\end{equation}
since congested traffic with zero flow and maximum density is upstream of the shock front.
\par
Obviously, the jam resolution has reached the further growing, upstream jam front when $C_it = c(t-t_0)$.
Therefore, the jam of density $\rho_{\rm max}$ has
disappeared after a time period $t-t_0 = C_it_0/(c - C_i)$, i.e. at time
\begin{equation}
t_1 = ct_0/(c - C_i) \,  .
\end{equation}
Surprisingly, it can be shown that $t_1 < t_2$, i.e. the traffic jam resolves before the number of delayed
vehicles reaches a value of zero. In fact, it still
takes the time $C_it_1/\widehat{V}_i^0$ until the last delayed vehicle has left the road section, where
\begin{equation}
 \widehat{V}_i^0 = \frac{A_i -O_i}{A_i/V_i^0 - (\rho_{\rm max} - O_i/c)}
\end{equation}
is the shock front between free upstream traffic flow and the congested outflow $O_i$,
which usually differs from the speed $V_i = O_i/[(1-TQ_i)\rho_{\rm max}]$ of outflowing vehicles.
For $O_i = \widehat{Q}_i$, we have $\widehat{V}_i^0 = V_i^0$ because of
$1/c = \rho_{\rm max}/\widehat{Q}_i - 1/V_i^0$.
\par
Undelayed traffic starts when this shock front reaches the end of the road section, i.e. at time
\begin{equation}
t_2 = t_1 \left(1 + \frac{C_i}{\widehat{V}_i^0}\right)
= \frac{t_0}{1 - C_i/c} \left( 1 + \frac{C_i(A_i/V_i^0-\rho_{\rm max}) + C_iO_i/c}{A_i - O_i} \right) \, .
\end{equation}
Inserting $C_i(A_i/V_i^0-\rho_{\rm max}) = - A_i$ eventually gives $t_2 = t_0 O_i/O_i - A_i)$.
This agrees perfectly with the above result for the first method (based on vehicle delays rather than traffic jams).
\par
In conclusion, both methods of dealing with traffic jams are consistent, and delayed vehicles occur as soon as traffic jam formation begins.
However, according to method 1, a queued vehicle at position $x_i = L_i - l_i$
is counted as delayed only after an extra time period
$l_i/V_i^0$, but it is counted as undelayed after the same extra time period. This is because method 1 counts on the basis of vehicle
arrivals at the downstream end of road section $i$.
\par
As it is much simpler to use the method 1 based on determining the number of delayed vehicles than using method 2 based on determining the
movement of shock fronts, we will use method 1 in the following. More specifically, in Eq.~(\ref{one}) we will replace $l_j(t) < L_j$ by
$\Delta N_j(t) < N_j^{\rm max} := L_j \rho_{\rm max}$ and $l_j(t) = L_j$ by
$\Delta N_j(t) = N_j^{\rm max}$. This corresponds to a situation in which the vehicles would not queue up along the
road section, but at the downstream end of the road section, like in a wide parking lot or on top of each other. As long as road section $j$
is not fully congested, this difference does not matter significantly. If it is fully congested, the dynamics will potentially be
different, defining a modified model of material network flows. However both, the original and the
modified model fulfill the conservation equation and show spillover effects.

\subsubsection{Calculation of Cumulative and Maximum Individual Waiting Times}

In Ref. \cite{JPhysA}, we have derived a delay differential equation to determine the travel time $T_i(t)$ of a vehicle entering road section
$i$ at time $t$ (see also Ref. \cite{Astarita1995,Astarita2002,Carey2003}):
\begin{equation}
 \frac{dT_i(t)}{dt} = \frac{A_i(t)}{O_i(t+T_i(t))} - 1 \, .
 \label{travtime}
\end{equation}
According to this, the travel time $T_i(t)$ increases with time, when the arrival rate $A_i$ at the time $t$ of entry exceeds the departure
rate $O_i$ at the leaving time $t+T_i(t)$, while it decreases when it is lower. It is remarkable that this formula does not explicitly depend
on the velocities on the road section, but only on the arrival and departure rates.
\par
Another method to determine the travel times is to integrate up over the number of vehicles arriving in road section $i$,
\begin{equation}
 N_i^A(t) = \int\limits_0^t dt' \; A_i(t') = N_i^{\rm in}(t+L_i/V_i^0) \, ,
\end{equation}
and over the number of vehicles leaving it,
\begin{equation}
 N_i^O(t) = \int\limits_0^t dt' \; O_i(t') = N_i^{\rm out}(t) \, ,
\end{equation}
starting at at time $t=0$ when there are no vehicles in the road. If $T'_i(t)$ denotes the time at which $N_i^O(t+T'_i(t)) = N_i^A(t)$, then
$T'_i(t)$ is the travel time of a vehicle entering road section $i$ at time $t$ and
\begin{equation}
 T_i(t) = T'_i(t) - L_i/V_i^0
\end{equation}
is its waiting time.
\par
Another interesting quantity is the cumulative waiting time $T_i^{\rm c}(t)$, which is determined by integrating up over the number $\Delta
N_i$ of all delayed vehicles. We obtain
\begin{eqnarray}
 T_i^{\rm c}(t) &=& \int\limits_0^t dt' \; \Delta N_i(t') = \int\limits_0^t dt'\; [N_i^{\rm in}(t'-L_i/V_i^0) - N_i^{\rm out}(t')]
\nonumber \\
&=& \int\limits_0^t dt' \int\limits_0^{t'} dt^{\prime\prime} \; [A_i(t^{\prime\prime}-L_i/V_i^0) - O_i(t^{\prime\prime}) ]
\end{eqnarray}
and the differential equation
\begin{equation}
 \frac{dT_i^{\rm c}(t)}{dt} =  \Delta N_i(t) = \int\limits_0^t dt' \; [A_i(t'-L_i/V_i^0) - O_i(t') ] \, .
\end{equation}
For a constant arrival flow $A_i$ and a red traffic light from $t=0$ to $t=t_0$ (i.e. $O_i(t) =0$), we find
\begin{equation}
 T_i^{\rm c} = \frac{A_i t^2}{2} \, .
\end{equation}
In this time period, a number of $N_i(t) = A_i t$ vehicles accumulates, which gives an average waiting time of
\begin{equation}
 \frac{T_i^{\rm c}(t_0)}{\Delta N(t_0)} = \frac{t_0}{2}
\end{equation}
at the end of the red light. The first vehicle has to wait twice as long, namely, a time period $t_0$.

\section{Treatment of Merging, Diverging and Intersection Points}\label{Sec3}

While the last section has given general formulas that must be fulfilled at nodes connecting two or more different links, in the following we
will give some concrete examples, how to deal with standard elements of street networks. For previous
treatments of traffic flows at intersections see, for example, Refs.~\cite{Piccoli1,Hilliges,CellTransmission,Lebacque}.
\begin{figure}[htbp]
\begin{center}
 \includegraphics[width=13cm, angle=0]{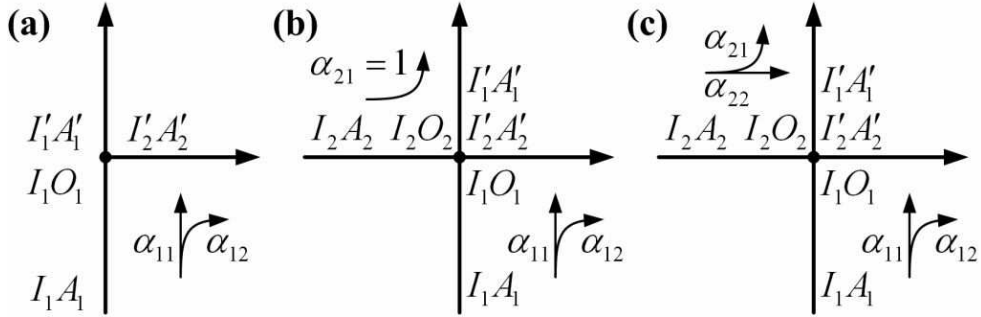}
\end{center}
\caption[]{Schematic illustration of the (a) diverging, (b) merging, and (c) intersecting flows
discussed in this paper.\label{Fig1}}
\end{figure}
\subsection{Diverging Flows: One Inflow and Several Outflows}

In the case of one road section $i$ diverging into several road sections $j$ (see Fig.~\ref{Fig1}a),
Eqs.~(\ref{nomerge}) and (\ref{easy1}) to
(\ref{inequal}) imply
\begin{eqnarray}
 O_i(t) &\le& \min \left\{ \gamma_i(t) \max \left[\widehat{Q}_i\Theta(\Delta N_i > 0), A_i\left(t-\frac{L_i}{V_i^0}\right)\right] ,\right.
\nonumber \\
& & \qquad\quad \left. \min_j \left[ \frac{I'_j}{I_i\alpha_{ij}} \max \left( \widehat{Q}_j \Theta(l_j < L_j),
 O_j(t- L_j / c) \right) \right] \right\}
\end{eqnarray}
for all $i$. If we assume that downstream road sections are never completely congested, this simplifies to
\begin{equation}
 O_i(t) = \min \left\{ Q_i, \gamma_i \max \left[\widehat{Q}_i\Theta(\Delta N_i > 0),
 A_i\left(t-L_i/V_i^0 \right)\right]
\right\}
\end{equation}
with
\begin{equation}
 Q_i = \min_j \left( \frac{I'_j\widehat{Q}_j}{I_i\alpha_{ij}} \right) \, .
\end{equation}
Otherwise
\begin{equation}
 Q_i(t) = \min_j \left[ \max\left( \frac{I'_j\widehat{Q}_j}{I_i\alpha_{ij}}\Theta(l_j < L_j) ,
\frac{I'_jO_j(t-\frac{L_j}{c})}{I_i\alpha_{ij}} \right)\right] \, . \label{otherwise}
\end{equation}

\subsection{Merging Flows: Two Inflows and One Outflow}

We assume a flow $I_1O_1(t)$ that splits into two flows $I_1O_1(t)\alpha_{11} $ (going straight) and $I_1O_1(t)\alpha_{12} $ (turning right),
but a right-turning flow $I_2O_2(t)$ merging with flow $I_1O_1(t) \alpha_{11}$, as in turn-right-on-red setups (see Fig.~\ref{Fig1}b). For
this situation, we have the equations
\begin{eqnarray}
 I'_1A'_1(t) &=& I_1O_1(t) \alpha_{11} + I_2O_2(t) \le I'_1\widehat{A}'_1(t) \, , \\
 I'_2A'_2(t) &=& I_1O_1(t) \alpha_{12} \le I'_2\widehat{A}'_2(t) \, .
\end{eqnarray}
One can derive
\begin{equation}
 0 \le O_1 = \min\Big[ \widehat{O}_1(t), \frac{I'_1\widehat{A}'_1(t)-I_2O_2(t)}{I_1\alpha_{11}},
\frac{I'_2\widehat{A}'_2(t)}{I_1\alpha_{12}}\Big]
\end{equation}
and
\begin{equation}
 0 \le O_2 = \min\Big[ \widehat{O}_2(t), \frac{I'_1\widehat{A}'_1(t)-I_1O_1(t)\alpha_{11}}{I_2} \Big] \, .
\end{equation}
Let us set
\begin{equation}
 O_1 = \min\Big[\widehat{O}_1, \frac{I'_1\widehat{A}'_1(t)}{I_1\alpha_{11}},
\frac{I'_2\widehat{A}'_2(t)}{I_1\alpha_{12}}\Big] \label{speci}
\end{equation}
and
\begin{equation}
 O_2(O_1) = \min\Big[ \widehat{O}_2(t), \frac{I'_1\widehat{A}'_1(t)-I_1O_1\alpha_{11}}{I_2} \Big] \, .
\label{maxi}
\end{equation}
Then, it can be shown that $O_2(t) \ge 0$ and $O_1(t) \le [I'_1\widehat{A}'_1(t)-I_2O_2(t)]/(I_1\alpha_{11})$, as demanded. If $O_1(t)$ is
chosen a value $\Delta O_1$ smaller than specified in Eq.~(\ref{speci}), but $O_2$ is still set to the maximum related value $O_2(O_1-\Delta
O_1)$ according to Eq.~(\ref{maxi}), the overall flow
\begin{equation}
 F=I_1O_1+I_2O_2
\end{equation}
is reduced as long as $\alpha_{11} < 1$, since this goes along with additional turning flows (while the number of lanes does not matter!).
Therefore, it is optimal to give priority to the outflow $O_1(t)$ according to Eq.~(\ref{speci}) and to add as much outflow $O_2(t)$ as
capacity allows. This requires suitable flow control measures, otherwise the optimum value of the overall flow $F$ could not be reached. In
fact, the merging flow would ``steel'' some of the capacity reserved for the ``main'' flow ($i=1$), which would reduce the possible outflow
$O_1(t)$ and potentially cause a breakdown of free traffic flow, as it is known from on-ramp areas of freeways \cite{mitTreiber} .

\subsection{A Side Road Merging with a Main Road}

Compared to the last section, the situation simplifies, if we have just a side road or secondary turning flow merging with a the flow of a
main road without any turning flow away from the main road. In this case, we have $\alpha_{11}=1$ and $\alpha_{12} =0$, which leaves us with
the relationships
\begin{equation}
 O_1 = \min\Big[\widehat{O}_1, \frac{I'_1\widehat{A}'_1(t)}{I_1}\Big]
\end{equation}
and
\begin{equation}
 O_2(O_1) = \min\Big[ \widehat{O}_2(t), \frac{I'_1\widehat{A}'_1(t)-I_1O_1}{I_2} \Big] \, .
\end{equation}
according to Eqs.~(\ref{speci}) and (\ref{maxi}).

\subsection{Intersection-Free Designs of Road Networks}

\begin{figure}[htbp]
\begin{center}
 \includegraphics[width=5.5cm, angle=0]{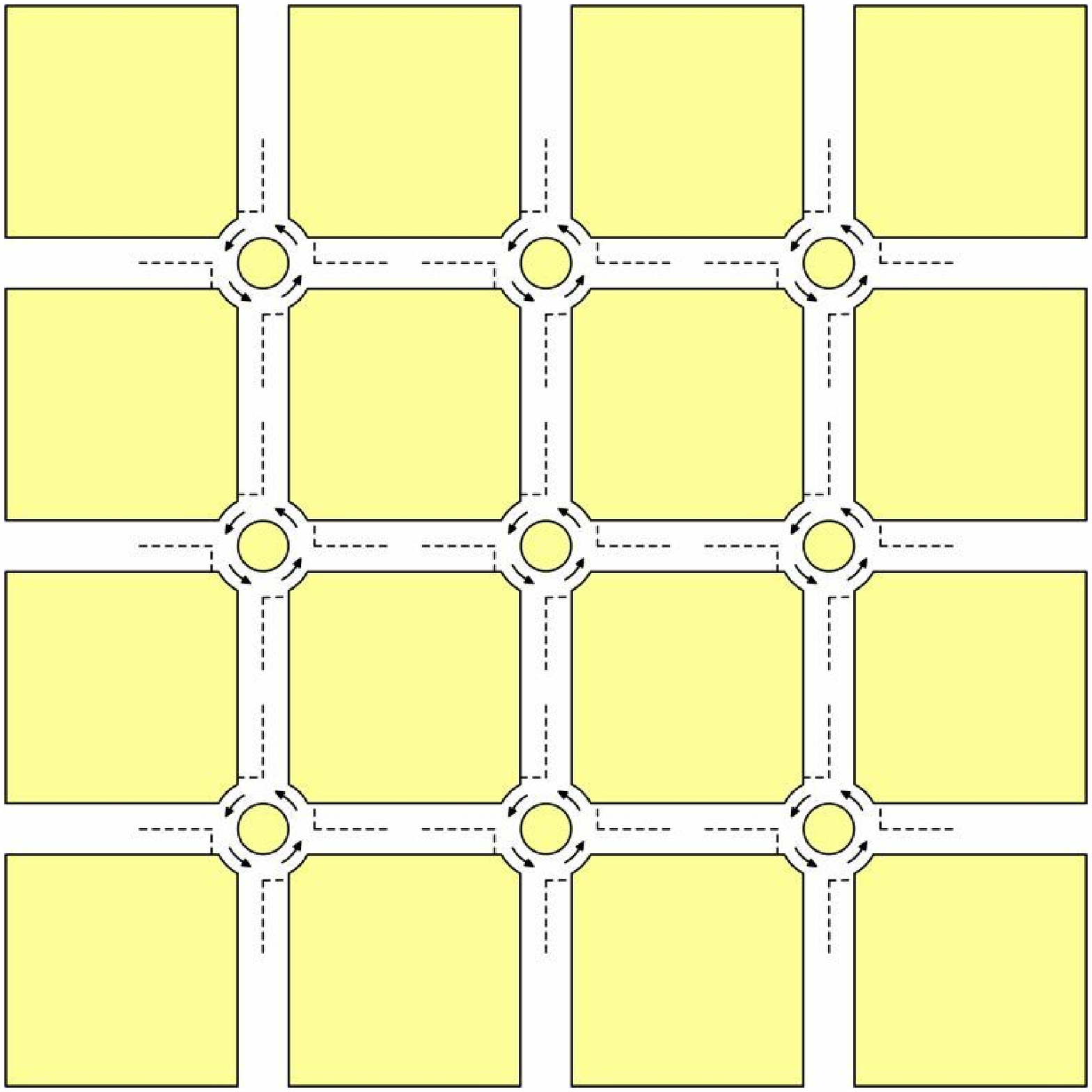}\hspace*{1cm}
 \includegraphics[width=6cm, angle=0]{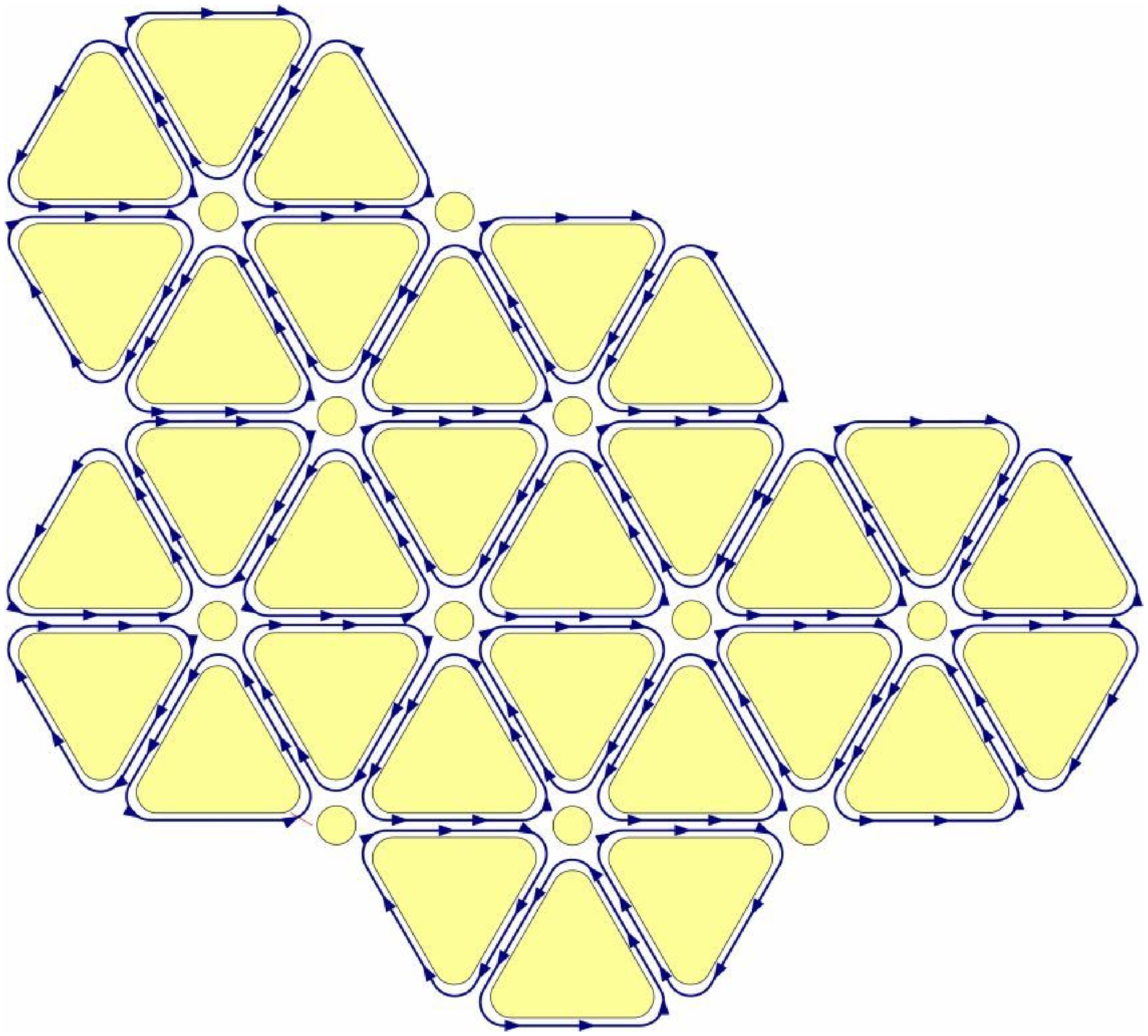}\\
 \includegraphics[width=6cm, angle=0]{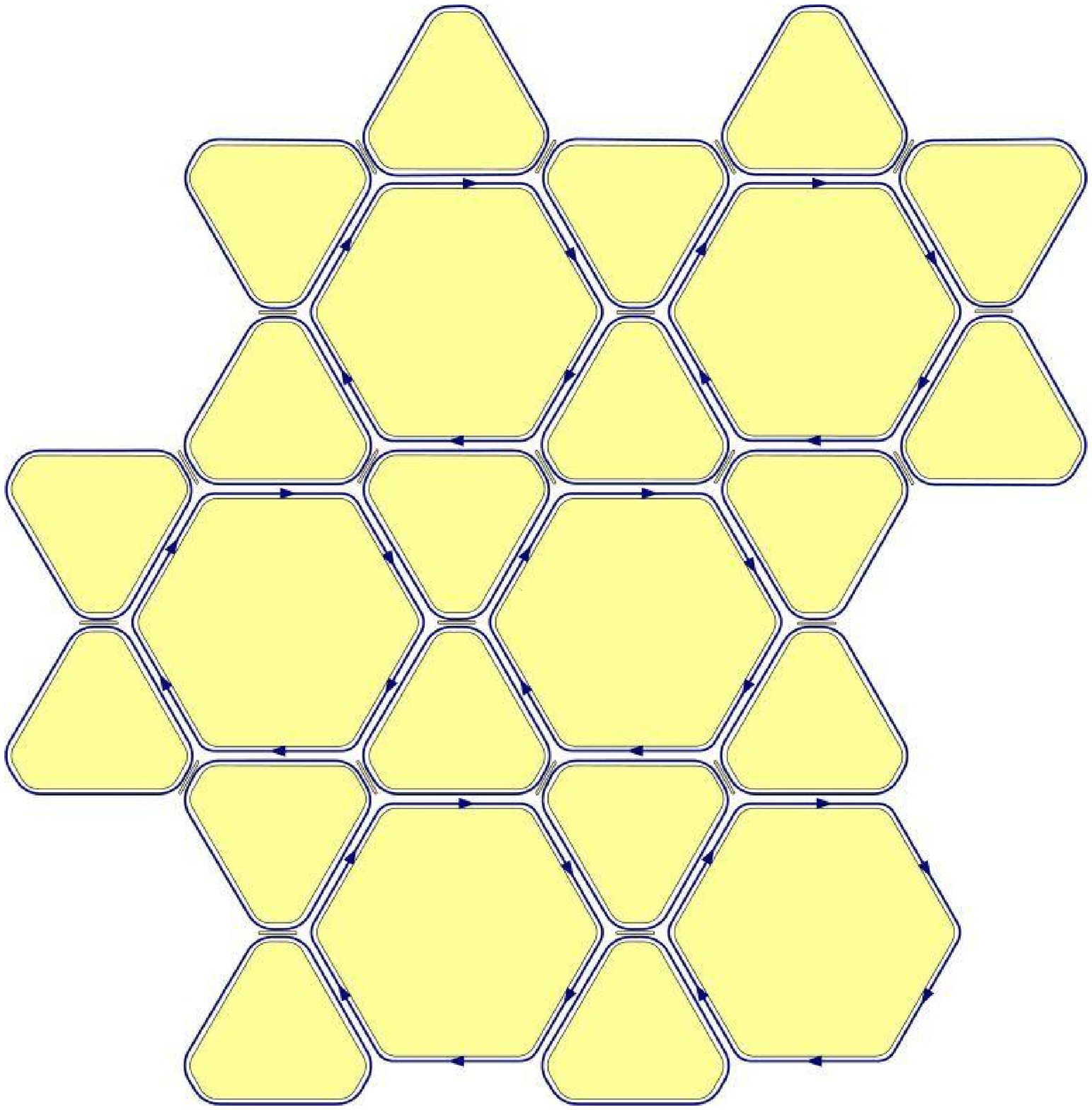}
\end{center}
\caption[]{Three examples for intersection-free designs of urban road networks.\label{free}}
\end{figure}

With the formulas for the treatment of merges and diverges in the previous sections, it is already possible
to simulate intersection-free designs of urban road networks, which do not need any traffic light control.
The most well-known design of intersection-free nodes are roundabouts (see the upper left illustration in Fig.~\ref{free}). It is,
however, also possible to construct other intersection-free designs based on subsequent merges and
diverges of flows with different destinations. Two examples are presented in Fig.~\ref{free}b and c.
\par
Although intersection-free designs require the driver to take small detours, such a road network will
normally save travel time and fuel, given that the traffic volume is not too low. This is because intersections then
need to be signalized in order to be safe and efficient.\footnote{Of course, a first-come-first-serve or right-before-left
rule will be sufficient at small traffic volumes.}
Traffic signals, however, imply that vehicles will often be stopped for considerable time intervals.
This causes significant delays, at least for vehicles not being served by a green wave. Intersection-free
designs, in contrast, do not necessarily require vehicles to stop. Therefore, the average speeds are expected to be
higher and the travel times lower than for road networks with intersections. This has significant
implications for urban transport planning, if intersections cannot be avoided by bridges or tunnels.

\subsection{Two Inflows and Two Outflows}

The treatment of intersecting flows is more complicated than the treatment of merges and diverges.
Moreover, the resulting flows are only uniquely defined, if additional rules are introduced such as
the optimization of the overall flow.
Let us here treat the case of an intersection with two inflows and two outflows (see Fig.~\ref{Fig1}c).
Equation (\ref{easy1}) implies the inequalities
\begin{eqnarray}
 & & 0 \le I'_1A'_1(t) = I_1O_1(t) \alpha_{11} + I_2O_2(t) \alpha_{21} \le I'_1 \widehat{A}'_1(t) \, , \nonumber \\
 & & 0 \le I'_2A'_2(t) = I_1O_1(t) \alpha_{12} + I_2O_2(t) \alpha_{22} \le I'_2 \widehat{A}'_2(t)
\label{constraints}
\end{eqnarray}
with the constraints
\begin{eqnarray}
& & 0 \le O_1(t) \le \widehat{O}_1(t) \, , \nonumber \\
& & 0 \le O_2(t) \le \widehat{O}_2(t) \, , \label{rectangle}
\end{eqnarray}
so that $I'_jA'_j(t) \ge 0$ is automatically fulfilled. The constraints (\ref{rectangle}) define an rectangular area of possible $O_i$-values
in the $O_1$-$O_2$ plane, where the size of the rectangle varies due to the time-dependence of $\widehat{O}_i(t)$. The inequalities
(\ref{constraints}) can be rewritten as
\begin{equation}
  O_2(t) \le  \frac{I'_1\widehat{A}_1(t) - I_1O_1(t)\alpha_{11}}{I_2\alpha_{21}}
 =: a_1 - b_1 O_1(t) \, , \label{const1}
\end{equation}
and
\begin{equation}
  O_2(t) \le  \frac{I'_2\widehat{A}_2(t) - I_1O_1(t)\alpha_{12}}{I_2\alpha_{22}}
 =: a_2 - b_2 O_1(t) \, . \label{const2}
\end{equation}
They potentially cut away parts of this rectangle, and the remaining part defines the convex set of feasible points $(O_1,O_2)$ at time $t$.
We are interested to identify the ``optimal'' solution $(O_1^*,O_2^*)$, which maximizes the overall flow
\begin{equation}
  \sum_j I'_j A'_j(t) = \sum_i I_i O_i(t) \, .
\end{equation}
As this defines a linear optimization problem, the optimal solution corresponds to one of the corners of the convex set of feasible points,
namely the one which is touched first by the line
\begin{equation}
 O_2 = \frac{Z - I_1 O_1}{I_2} \, ,
\label{goal1}
\end{equation}
when we reduce $Z$ from high to low values.
\par
Let us, therefore, determine all possible corners of the convex set and the conditions, under which they correspond to the optimal solution.
We will distinguish the following cases:
\begin{itemize}

\item[(a)] {\it None} of the boundary lines (\ref{const1}) and (\ref{const2}) corresponding to the equality signs
cuts the rectangle defined by $0 \le O_1(t) \le \widehat{O}_1(t)$ and $0 \le O_2(t) \le \widehat{O}_2(t)$ in more than 1 point. This case
applies, if $a_1 - b_1\widehat{O}_1 \ge \widehat{O}_2$ and $a_2 - b_2\widehat{O}_1 \ge \widehat{O}_2$, as $a_i\ge 0$ and $b_i \ge 0$ implies
that both lines are falling or at least not increasing. Since the line (\ref{goal1}) reflecting the goal function is falling as well, the
optimal point is
\begin{equation}
  (O_1^*,O_2^*) = (\widehat{O}_1,\widehat{O}_2) \, ,
\end{equation}
i.e. the outer corner of the rectangle corresponding to the potential or maximum possible departure flows (see Fig.~\ref{Fig2}).
\begin{figure}[htbp]
\begin{center}
  \includegraphics[width=6cm, angle=0]{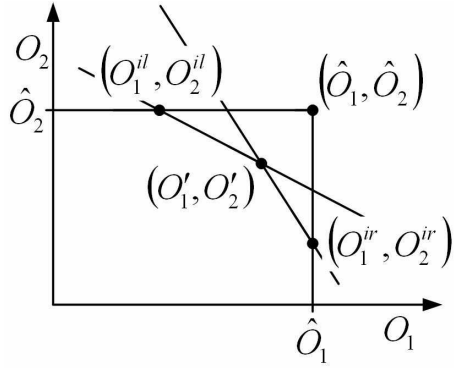}
\end{center}
\caption[]{Illustration of the possible optimal solutions for two intersecting flows (see text for details).\label{Fig2}}
\end{figure}
\item[(b)] Only {\it one} of the two boundary lines/border lines,
$O_2(t) = a_1-b_1O_1(t)$ or $O_2(t) = a_2 - b_2O_1(t)$, cuts the rectangle in {\it more} than one point. Let us assume, this holds for line
$i$, i.e. $a_i - b_i\widehat{O}_1 < \widehat{O}_2$. Then, the left cutting point
\begin{equation}
\qquad (O_1^{i{\rm l}},O_2^{i{\rm l}}) = \left\{
\begin{array}{ll}
\Big((a_i - \widehat{O}_2)/b_i,\widehat{O}_2\Big) & \mbox{if } a_i > \widehat{O}_2 \, ,\\
(0,a_i) & \mbox{otherwise}
\end{array}\right.
\end{equation}
is the optimal point if $I_1/I_2 < b_i$, i.e. if the slope $I_1/I_2$ of the goal function (\ref{goal1}) is smaller than the one of the
cutting border line. Otherwise, if $I_1/I_2 > b_i$, the optimal point is given by the right cutting point
\begin{equation}
\qquad (O_1^{i{\rm r}},O_2^{i{\rm r}}) = \left\{
\begin{array}{ll}
 (\widehat{O}_1,a_i-b_i\widehat{O}_1) &  \mbox{if } a_i > b_i\widehat{O}_1 \, , \\
(a_i/b_i,0) & \mbox{otherwise}
\end{array}\right.
\end{equation}
(see Fig.~\ref{Fig2}).

\item[(c)] If {\it both} border lines cut through the rectangle, but one of them lies above the other line,
then only the lower line determines the optimal solution, which can be obtained as in case (b). Case (c) occurs if $a_2-b_2O_1^{1{\rm l}} >
a_1-b_1O_1^{1{\rm l}}$ and $a_2-b_2O_1^{1{\rm r}} >  a_1-b_1O_1^{1{\rm r}}$ (line 1 is the lower one) or if $a_2-b_2O_1^{1{\rm l}} <
a_1-b_1O_1^{1{\rm l}}$ and $a_2-b_2O_1^{1{\rm r}} <  a_1-b_1O_1^{1{\rm r}}$ (line 2 is the lower one).

\item[(d)] The boundary lines cut each other in the inner part of the rectangle. This occurs if
$a_1 - b_1\widehat{O}_1 < \widehat{O}_2$ and $a_2 - b_2\widehat{O}_1 < \widehat{O}_2$. Then, the left-most cutting point $ (O_1^{i{\rm
l}},O_2^{i{\rm l}})$ is the optimal solution, if the slope $I_1/I_2$ of the goal function is smaller than the smallest slope of the two
boundary lines, while it is the lower right cutting point $ (O_1^{i{\rm r}},O_2^{i{\rm r}})$, if $I_1/I_2$ is greater than the steepest slope
of the two boundary lines, otherwise, the cutting point of the two boundary lines,
\begin{equation}
 (O'_1,O'_2) = \left( \frac{a_2 - a_1}{b_2-b_1} , \frac{a_1b_2 - b_1a_2}{b_2-b_1} \right)
\end{equation}
is the optimal point (see Fig.~\ref{Fig2}). Mathematically speaking, we have
\begin{equation}
 (O_1^*,O_2^*) = \left\{
\begin{array}{ll}
 (O_1^{\rm 1l},O_2^{\rm 1l}) & \mbox{if } I_1/I_2 < b_1 < b_2, \\
 (O'_1,O'_2) & \mbox{if } b_1 < I_1/I_2 < b_2, \\
 (O_1^{\rm 2r},O_2^{\rm 2r}) & \mbox{if } b_1 < b_2 < I_1/I_2, \\
 (O_1^{\rm 2l},O_2^{\rm 2l}) & \mbox{if } I_1/I_2 < b_2 < b_1, \\
 (O'_1,O'_2) & \mbox{if } b_2 < I_1/I_2 < b_1, \\
 (O_1^{\rm 1r},O_2^{\rm 1r}) & \mbox{if } b_2 < b_1 < I_1/I_2,
\end{array}\right.
\end{equation}
\end{itemize}
\par
It is astonishing that the simple problem of two intersecting traffic flows has so many different optimal solutions, which sensitively depend
on the parameter values. This can reach from situations where both outgoing road sections experience the maximum possible outflows upto
situations, where the outflow in the system-optimal point becomes zero for one of the road sections. A transition from one optimal solution
to another one could easily be triggered by changes in the turning fractions $\alpha_{ij}$ entering the parameters $a_i$ and $b_i$, for
example due to time-dependent turning fractions $\alpha_{ij}(t)$.

\subsection{Inefficiencies due to Coordination Problems}

An interesting question is how to actually establish the flows corresponding to the system optima that were determined in the previous sections on
merging and intersecting flows. Of course, zero flows can be enforced by a red traffic light, while maximum possible flows can be established
by a node design giving the right of way to one road (the ``main'' road). However, it is not so easy to support an optimimum point
corresponding to mixed flows, such as $(O'_1,O'_2)$. That would need quite tricky intersection designs or the implementation of an
intelligent transportation system ensuring optimal gap usage, e.g. based on intervehicle communication. Only in special cases, the task could
be performed by a suitable traffic light control.
\par
In normal merging or intersection situations, there will always be coordination problems \cite{mitJohansson} when entering or crossing
another flow, if the traffic volumes reach a certain level. This will cause inefficiencies in the usage of available road capacity, i.e. mixed flows will not
be able to use the full capacity. Such effects can be modelled by specifying the corresponding permeabilities $\gamma_i(t)$ as a function of
the merging flows, particularly the main flow or crossing flow. The deviation of $\gamma_i(t)$ from 1 will then be a measure for the
inefficiency. A particularly simple, phenomenological specification would be
\begin{equation}
\gamma_2(t) =
\frac{1}{1+ a\mbox{e}^{b(O_1-O_2)}} \, ,
\label{phen}
\end{equation}
where the own outflow $O_2$ supports a high permeability and the intersecting outflow $O_1$ suppresses it.
However, rather than using such a phenomenological approach, the permeability could also be calculated analytically, based on
a model of gap statistics, since large enough vehicle gaps are needed to join or cross a flow. Such kinds of calculations have been carried
out in Refs.~\cite{analyticJiang,Biometrica,Troutbeck1986,Troutbeck1997}.
\section{Towards a Self-Organized Traffic Light Control}\label{Sec4}

In Ref.~\cite{control}, it has been pointed out that, for not too small arrival flows, an oscillatory service at intersections reaches higher
intersection capacities and potentially shorter waiting times than a first-in-first-out service of arriving vehicles. This is due to the fact that
the outflow of queued vehicles is more efficient than waiting for the arrival of other freely flowing vehicles, which have larger time gaps. For
similar reasons, pedestrians are passing a bottleneck in an oscillatory way \cite{mitMolnar}, and also two intersecting flows tend to
organize themselves in an oscillatory way \cite{analyticJiang,TransportationScience}.
\par
Therefore, using traffic lights at intersections is natural and useful, if operated in the right way. However, how to switch the traffic lights
optimally? While this is a solvable problem for single traffic lights, the optimal coordination of many traffic lights
\cite{Papadimitriou1999} is a really hard (actually NP hard) problem \cite{Schutter2002}. Rather than solving a combinatorial optimization problem, here, we
want to suggest a novel approach, which needs further elaboration in the future. The idea is to let the network flows self-organize
themselves, based on suitable equations for the permeabilities $\gamma_i(t)$ as a function of the outflows $O_i(t)$ and the numbers $\Delta
N_i(t)$ of delayed vehicles.
\par
\begin{figure}[htbp]
\begin{center}
  \includegraphics[width=\textwidth, angle=0]{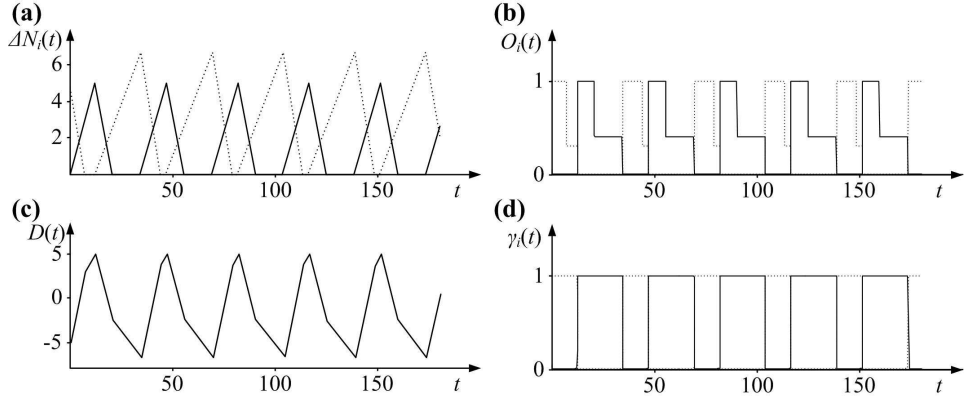}
\end{center}
\caption[]{Illustration of the dynamics of self-organized oscillations in the permeabilities and
the resulting flows for a single intersection with constant inflows (see text for details). Note that
the road section with the higher inflow (arrival rate) is served longer, and
its queues are shorter (see solid lines).\label{ill}}
\end{figure}

Here, we will study the specification
\begin{equation}
\gamma_1(t) = \frac{1}{1+ a\mbox{e}^{b(O_2-O_1) - cD}}
\label{pre1}
\end{equation}
and
\begin{equation}
\gamma_2(t) = \frac{1}{1+ a\mbox{e}^{b(O_1-O_2) + cD}} \, ,
\label{pre2}
\end{equation}
which generalizes formula (\ref{phen}). While the relative queue length
\begin{equation}
D(t)=\Delta N_1(t) - \Delta N_2(t)
\end{equation}
quantifies the pressure to increase the permeability $\gamma_1$ for road section 1,
the outflow $O_2(t)$ from the road section 2 resists this tendency, while the flow
$O_1(t)$ on road section 1 supports the permeability.
The increasing pressure eventually changes the resistance threshold and
the service priority. An analogous situation applies to the permeability $\gamma_2$ for road section 2,
where the pressure corresponds to $-D$, which is again the difference in queue length.
$a$, $b$, and $c$ are non-negative parameters. $a$ may be set to 1, while $c$ must be large enough
to establish a sharp switching. Here, we have assumed $c=100$.
The parameter $b$ allows to influence the switching frequency $f$,
which is approximately proportional to $b$. We have adjusted the frequency $f$ to the cycle time
\begin{equation}
 T^{\rm cyc} = \frac{2\tau}{1-(A_1+A_2)/\widehat{Q}} \, ,
\label{cycl}
\end{equation}
which results if the switching (setup) time (``yellow traffic light'') is $\tau = 5$s and
a green light is terminated immediately after a queue has dissolved after lifting the red
light.\footnote{If $\Delta T_1$ and $\Delta T_2$ denote the green time periods for the intersecting
flows 1 and 2, respectively, the corresponding red time periods for a periodic signal control
are $\Delta T_2$ and $\Delta T_1$, to which
the switching setup time of duration $\tau$ must be added. From formula (\ref{acco}) and with
$O_i = \widehat{Q}$ we obtain $\Delta T_1 = (\Delta T_2 +\tau)\widehat{Q}/(\widehat{Q}-A_1)$
and $\Delta T_2 = (\Delta T_1 +\tau)\widehat{Q}/(\widehat{Q}-A_2)$. Using the definition
$T^{\rm cyc} = \Delta T_1 + \tau + \Delta T_2 + \tau$ for the cycle time, we finally arrive at Eq.~(\ref{cycl}).}
The corresponding parameter value is
\begin{equation}
 b = \frac{500}{\widehat{Q} - (A_1+A_2)} \, .
\label{choice}
\end{equation}
Figure \ref{ill} shows a simulation result for $A_1/\widehat{Q} = 0.3$ and $A_2/\widehat{Q} = 0.4$.
\par
The properties of the corresponding specification of the permeabilitities $\gamma_i(t)$ are as follows:
\begin{itemize}
\item $\gamma_i(t)$ is non-negative and does not exceed the value 1.
\item For the sum of permeabilities and $a\ge 1$, we have
\begin{equation}
\gamma_1 +\gamma_2  = \frac{2+a(\mbox{e}^E + \mbox{e}^{-E})}{1+ a^2 + a(\mbox{e}^E + \mbox{e}^{-E})}
\le 1 \, ,
\end{equation}
where we have introduced the abbreviation
\begin{equation}
 E = b(O_1 - O_2) + c(\Delta N_1 - \Delta N_2) \, .
\end{equation}
The sum is close to 1 for large absolute values of $E$, while for $E \approx 0$
the overall permeability $\gamma_1 + \gamma_2$ is small.
\item For large enough values of $ab$ and for $c, A_1, A_2 >0$,
the equations for the permeability do not have a stable stationary solution.
This can be concluded from
\begin{equation}
 \frac{dE}{dt} = b \left( \frac{dO_1}{dt} - \frac{O_2}{dt}\right)
 + c \left( \frac{d\Delta N_1}{dt} - \frac{d\Delta N_2}{dt}\right)
\end{equation}
together with
\begin{equation}
 \frac{d\Delta N_i}{dt} = A_i - O_i(t)
\end{equation}
and
\begin{equation}
 O_i(t) = \gamma_i(t) \max[\widehat{Q}\Theta(\Delta N_i> 0), A_i] \, ,
\end{equation}
see Eqs. (\ref{delayed}) and (\ref{easy2}). As
$dD/dt = d\Delta N_1/dt - d\Delta N_2/dt$ varies around
zero, the same applies to $D(t)$, which leads to oscillations of the permeabilities $\gamma_i(t)$.
\item With the specification (\ref{choice}) of parameter $b$,
the cycle time is approximately proportional to the overall inflow $(A_1+A_2)$.
\item The road section with the higher flow gets a longer green time period
(see Fig. \ref{ill}).
\end{itemize}
If the above self-organized traffic flows shall
be transfered to a new principle of traffic light control, phases with $\gamma_i(t) \approx 1$ could be
interpreted as green phases and phases with $\gamma_i(t)\approx 0$ as red phases. Inefficient,
intermediate switching time periods for certain choices of parameter values could be translated
into periods of a yellow traffic light.

\section{Summary and Outlook}\label{Sec5}

We have presented a simple model for conserved flows in networks. Although our specification has been illustrated for traffic flows in urban
areas, similar models are useful for logistic and production system or even transport in biological cells or bodies. Our model considers
propagation speeds of entities and congestion fronts, spill-over effects, and load-dependent transportation times.
\par
We have also formulated constraints for network nodes. These constraints contain several minimum and maximum functions, which implies a
multitude of possible cases even for relatively simple intersections. It turns out that the arrival and departure flows of diverges have
uniquely defined values, while merges or intersections have a set of feasible solutions. This means, the actual result may sensitively
depend on the intersection design. For mathematical reasons, we have determined flow-optimizing solutions for two merging and two intersecting
flows. However, it is questionable whether these solutions can be established in reality without the implementation of intelligent transport
systems facilitating optimal gap usage: In many situations, coordination problems between vehicles in merging or intersection areas cause
inefficiencies, which reduce their permeability.
\par
In fact, at not too small traffic volumes, it is better to have an oscillation between minimum and maximum permeability values. Therefore, we
have been looking for a mechanism producing emergent oscillations between high and low values. According to our proposed specification (which
is certainly only one of many possible ones), the transition between high and low permeability was triggered, when the difference between the
queue lengths of two traffic flows competing for the intersection capacity exceeded a certain value. The resulting oscillatory service could
be used to {\it define} traffic phases. One potential advantage of such an approach would be that the corresponding traffic light control would be based on
the self-organized dynamics of the system. Further work in this direction seems very promising.

\section*{Acknowledgements}

The authors are grateful for partial financial support by the German Research Foundation (research projects He 2789/5-1, 
8-1) and by the ``Cooperative Center
for Communication Networks Data Analysis'', a NAP project sponsored by
the Hungarian National Office of Research and Technology under grant
No.\ KCKHA005.

%
%
%
\end{document}